\documentclass[%
 reprint,
 amsmath,amssymb,
 aps,
 pra,
 superscriptaddress]{revtex4-2}

\usepackage[colorlinks=true, allcolors=blue]{hyperref}
\usepackage{amsmath}
\usepackage{braket}
\usepackage{graphicx}
\usepackage{dcolumn}
\usepackage{bm}

\begin{document}

\preprint{APS/123-QED}

\title{Hardware-Efficient Exchange-Only QML: Singlet-Triplet Spin Chains via Inter-pair Coupling without Magnetic Gradients}

\author{Yuichiro Minato}
\email{minato@blueqat.co.jp}
\affiliation{blueqat Inc., 2-24-12 Shibuya, Shibuya-ku, Tokyo 150-6139, Japan}

\date{\today}

\begin{abstract}
Universal quantum computing with exchange-only qubits typically requires three physical spins per logical qubit, adding hardware overhead. Two-spin units are denser but rely on local magnetic field gradients for control, which complicates integration. In this paper, we study a quantum machine learning (QML) architecture that uses two-spin units and Heisenberg exchange interactions alone, without any magnetic gradients.

Instead of universal gate-based control, we use the time-domain dynamics of a spin chain as a learning resource. Numerical simulations on MNIST digit classification show that the symmetry constraints of isolated spin pairs are relaxed by inter-pair exchange coupling, which mixes the singlet and triplet components and enlarges the reachable state space. The model reaches a test-set accuracy of $90.9\%$ on the full 10{,}000-image MNIST test set ($91.0\% \pm 0.2\%$ over five independent seeds); a hyperparameter search on a validation split held out from the training data identifies a larger configuration that reaches $93.3\%$. Under an identical linear readout, the trained model ($88.1\%$) outperforms a classical linear baseline on the same PCA inputs ($83.2\%$) and an untrained (reservoir-style) version of the same dynamics ($53.0\%$), so the pulse areas have to be learned. We also quantify what the chain itself contributes: the encoder applies a $\tanh$ non-linearity to the data before any quantum evolution, and controls in which the chain is replaced by a fixed, untrained quadratic map of the pulse areas reach comparable or higher accuracy, so we do not attribute the non-linearity of the feature map to the exchange dynamics. The eleven observables are native readout primitives of the platform, and the five pair correlators alone---each a singlet-probability measurement accessible by Pauli spin blockade---already give $89.4\%$, so the protocol does not require single-spin readout. The protocol tolerates experimentally relevant imperfections: accuracy remains at $89.9\%$ under $10\%$ quasi-static pulse-area noise and at $89.8\%$ when every observable is estimated from $10^3$ measurement shots. These results suggest that QML of this kind can run on minimal semiconductor spin-chain hardware, without leakage-prone encodings or micro-magnet integration.
\end{abstract}

\maketitle

\section{Introduction}

Semiconductor spin qubits in silicon-based quantum dots have emerged as a leading platform for scalable quantum information processing, owing to their long coherence times and compatibility with industrial CMOS fabrication processes \cite{Loss1998, Burkard2023}. While experimental progress toward large-scale integrated quantum systems has been substantial, the trade-off between logical qubit density and the control infrastructure remains a bottleneck for practical implementation \cite{Zajac2016, Taylor2005}.

In the development of spin-based control strategies, several approaches have been widely explored. One prominent direction involves the \textbf{three-spin exchange-only (EO) qubit} \cite{DiVincenzo2000}. While this architecture enables all-electrical control via the Heisenberg exchange interaction, it is inherently resource-inefficient, requiring three physical spins per logical qubit. Furthermore, executing universal gates within this decoherence-free subsystem frequently leads to transitions into unused sectors of the Hilbert space, known as leakage, which requires sophisticated pulse sequences to mitigate \cite{Fong2011}.

Another significant approach is the \textbf{two-spin singlet-triplet (S-T) qubit} \cite{Petta2005}. Although this offers a more compact footprint, universal control conventionally necessitates a local magnetic field gradient ($\Delta B_z$), typically generated by micro-magnets \cite{Brunner2011} or nuclear spin fields \cite{Hensen2020}. This requirement for heterogeneous hardware components complicates the fabrication process and serves as a potential source of magnetic noise, hindering the realization of a monolithic, all-electrical spin processor.

One way to relax these constraints is offered by \textbf{Variational Quantum Algorithms (VQAs)} \cite{Cerezo2021}. This framework prioritizes the optimization of variational parameters over rigid gate-sets, offering a flexible and hardware-efficient approach to quantum computation. By interpreting natural spin dynamics as a mapping into a feature Hilbert space \cite{Schuld2019, Havlicek2019}, VQAs allow for the re-evaluation of traditional logical encodings in favor of resource-optimized machine learning frameworks that are better suited for near-term quantum hardware.

\textit{Related work.}---The idea of using natural Hamiltonian dynamics as a computational resource has been explored in quantum reservoir computing (QRC) and quantum extreme learning machines (QELMs) \cite{Mujal2021}, including image classification with fixed Ising or spin-network reservoirs \cite{Sakurai2025, DeLorenzis2025} and large-scale analog implementations on neutral-atom hardware \cite{Kornjaca2024}. In these approaches, however, the quantum dynamics are \emph{fixed}, and only a classical readout is trained. A complementary line of work optimizes control pulses directly at the hardware level (pulse-based QML) \cite{Tao2025}, typically targeting superconducting platforms with drive fields. Our approach differs in two respects: (i) the data-dependent pulse areas of every exchange link are themselves trainable, making the protocol an analog counterpart of data re-uploading circuits \cite{PerezSalinas2020} rather than a static reservoir---an ablation with frozen random pulses (Sec.~\ref{sec:baselines}) shows this training is essential---and (ii) the control alphabet is restricted to the isotropic Heisenberg exchange interaction alone, the native, all-electrical resource of semiconductor spin chains, with no magnetic field gradients or ac drives. The $M_z$-conserving structure of this ansatz also connects to symmetry-equivariant strategies for mitigating barren plateaus \cite{McClean2018, Larocca2025, Meyer2023}, and, more broadly, to dynamics-based models whose trainability is controlled by the physical phase of the underlying spin chain, such as many-body-localized hidden Born machines \cite{Zhong2024}.

In this work, we study a Quantum Machine Learning (QML) architecture that uses a linear chain of two-spin units controlled solely by Heisenberg exchange interactions. We show numerically that the requirement for $\Delta B_z$ can be bypassed for this task by using the non-commutativity of overlapping \textbf{inter-pair exchange pulses} ($[H_{\text{intra}}, H_{\text{inter}}] \neq 0$), which provides enough state mixing for the classification problem studied here without any magnetic gradient.

Applied to MNIST digit classification, the model reaches $90.9\%$ test accuracy with a standard mini-batch training procedure. The time-domain dynamics of a Heisenberg spin chain, driven by exchange interactions alone, thus map PCA-reduced image features into distinguishable quantum states. This ``Time-Encoded Exchange'' (TE-EX) mechanism enlarges the reachable state space through the non-commutativity of the control pulses and offers a hardware-efficient way of running quantum machine learning tasks.

These results suggest that such learning tasks can be run on minimal semiconductor spin-chain hardware, without the leakage-prone encodings of three-spin exchange-only systems or the heterogeneous hardware of singlet-triplet qubits. More broadly, this work is a step toward using natural spin interactions as a computational resource in all-electrical spin-chain devices.

\section{Methodology}

\subsection{Physical System and Hamiltonian Dynamics}
We model a physical system consisting of a one-dimensional array of $N=6$ semiconductor quantum dots. Each dot is assumed to be in the single-electron regime, where the logical information is encoded in the electron spin degrees of freedom. The primary interaction between adjacent spins is the isotropic Heisenberg exchange interaction, which can be dynamically controlled via electrical gate voltages. The system is governed by the time-dependent Hamiltonian:
\begin{equation}
    H(t) = \sum_{i=1}^{N-1} J_i(t) (\mathbf{S}_i \cdot \mathbf{S}_{i+1}),
\end{equation}
where $\mathbf{S}_i = \frac{1}{2}(X_i, Y_i, Z_i)$ represents the spin operator at site $i$ defined by the Pauli matrices. The exchange term is given by the scalar product $\mathbf{S}_i \cdot \mathbf{S}_{i+1} = \frac{1}{4}(X_i X_{i+1} + Y_i Y_{i+1} + Z_i Z_{i+1})$. Throughout the numerical model we work with the dimensionless link generator
\begin{equation}
    h_{i,i+1} \equiv X_i X_{i+1} + Y_i Y_{i+1} + Z_i Z_{i+1} = 4\, \mathbf{S}_i \cdot \mathbf{S}_{i+1},
    \label{eq:generator}
\end{equation}
so that all pulse areas quoted below are expressed in units of $h_{i,i+1}$; multiplying by four converts them to units of $\mathbf{S}_i \cdot \mathbf{S}_{i+1}$. 

A key challenge in gradient-free architectures is that the isotropic exchange interaction alone, when acting on isolated spin pairs, is SU(2) symmetric and cannot induce transitions between the singlet $|S\rangle$ and triplet $|T_0\rangle$ states. In conventional singlet-triplet (S-T) qubit operations, this symmetry is typically broken by introducing a local magnetic field gradient ($\Delta B_z$), which enables universal control but significantly complicates hardware integration due to the need for micro-magnets or nuclear spin polarization.

To circumvent this requirement, our Time-Encoded Exchange (TE-EX) protocol exploits the \textbf{non-commutativity} of adjacent exchange interactions. Since the Hamiltonians of overlapping pairs do not commute, i.e., $[H_{i,i+1}, H_{i+1,i+2}] \neq 0$, a sequential application of exchange pulses across the chain mixes the singlet and triplet components of each pair. Mathematically, the total unitary evolution within a layer, $U = \prod e^{-i H_{i,i+1} \Delta t}$, cannot be simplified into a product of independent pair rotations. This mixing lets the system reach a larger set of states than isolated pairs can, and for the purpose of the feature map it plays the role that the magnetic gradient plays in conventional S-T control, without any external magnetic infrastructure.

\subsection{Initial State and Symmetry Constraints}
To ensure the numerical stability and physical realizability of the learning process, the system is initialized within a specific symmetry-protected subspace. We restrict the quantum dynamics to the zero-magnetic-projection sector ($M_z = \sum S_{z,i} = 0$), where the number of spin-up and spin-down electrons is equal. In our numerical implementation, the system is initialized in the computational basis state $|\Psi(0)\rangle = |010101\rangle$:
\begin{equation}
    |\Psi(0)\rangle = |01\rangle_{1,2} \otimes |01\rangle_{3,4} \otimes |01\rangle_{5,6},
\end{equation}
where $|0\rangle$ and $|1\rangle$ denote the spin-up and spin-down states, respectively. 

Physically, this Néel-like product state can be prepared without any magnetic field gradient: in a global uniform Zeeman field, each dot is initialized individually by energy-selective spin filling \cite{Elzerman2004}. A global field is compatible with the gradient-free premise of this work---it commutes with the Heisenberg Hamiltonian and merely shifts the $M_z$ sectors in energy.

The choice of initial state matters more than it may appear, because the isotropic exchange interaction conserves not only $M_z$ but also the total spin $S_{\mathrm{tot}}$. The seemingly natural alternative---a product of two-spin singlets $|S\rangle^{\otimes 3}$ prepared by adiabatic $(0,2) \to (1,1)$ loading \cite{Petta2005}---is a total-spin singlet, so exchange-only dynamics launched from it remains confined to the five-dimensional $S_{\mathrm{tot}}=0$ block of the $M_z = 0$ sector. The Néel state, in contrast, has support on all four total-spin blocks (of dimensions $5+9+5+1 = 20$), with weights exactly proportional to the block dimensions ($1/4$, $9/20$, $1/4$, $1/20$), so the trained dynamics acts nontrivially in every block. Because $[H, \mathbf{S}_{\mathrm{tot}}^2] = 0$, the weight in each $S_{\mathrm{tot}}$ sector is separately conserved: the accessible set is the orbit with these fixed sector weights rather than the entire 20-dimensional sector, but it is still a strictly larger arena than the single five-dimensional block reachable from the singlet product. Counting real parameters, a normalized state of the 20-dimensional sector has $2 \times 20 - 2 = 38$ real degrees of freedom once normalization and the global phase are removed; the four sector weights sum to one, so their conservation imposes three further independent constraints, and the states reachable by exchange from the Néel state lie in a manifold of real dimension at most $35$. This difference is quantitative: retraining the model with a singlet-product initial state under otherwise identical settings reaches $87.0\%$ test accuracy---the protocol still functions, but with reduced expressibility---so the Néel initialization is both the more expressive and the experimentally convenient choice.

By confining the time evolution to the $M_z=0$ sector, the effective Hilbert space dimension is reduced from $2^N = 64$ to $\binom{6}{3} = 20$. Since the Heisenberg Hamiltonian commutes with the total magnetization operator $[H, \sum S_{z,i}] = 0$, the system cannot leak into other $M_z$ sectors. This symmetry constraint reduces the simulation cost and restricts the optimization to the sector in which the singlet-triplet mixing takes place. It may also aid trainability at larger system sizes: ans\"atze whose dynamical Lie algebra is polynomially restricted by symmetry are known to avoid exponentially vanishing gradients \cite{Larocca2025}, and the exchange-only generators studied here are of exactly this symmetry-restricted type. We emphasize that at $N=6$ barren plateaus do not yet manifest, so we make no empirical claim about their mitigation.

\subsection{Reachable Transformations and Finite-Depth Expressivity}
\label{sec:reachable}
Since exchange conserves both $M_z$ and $\mathbf{S}_{\mathrm{tot}}^2$, the transformations available inside the $M_z = 0$ sector are block diagonal in the total-spin basis, and the group generated by the exchange pulses is contained in $SU(5) \times SU(9) \times SU(5) \times U(1)$ (the blocks of dimension $5 + 9 + 5 + 1 = 20$, with a single overall phase), of real dimension $24 + 80 + 24 + 1 = 129$. Closing the algebra generated by $\{h_{12}, \dots, h_{56}\}$ numerically under commutators---repeatedly adding commutators that are linearly independent of those already collected---terminates at exactly 129, with every element block diagonal in the $\mathbf{S}_{\mathrm{tot}}^2$ eigenbasis to within $10^{-7}$. The five nearest-neighbour exchange generators alone therefore span the entire symmetry-allowed algebra: nothing beyond the conservation laws restricts what the chain can do.

This is a statement about arbitrary control of the generator coefficients. The ansatz instead repeats the same five generators layer after layer and relies on the layered structure to emulate such control. A depth-$L$ ansatz has only $5L$ parameters, and in the trained model the pulse areas are further bounded by $|\theta| \le \gamma$ through the $\tanh$ parameterization. To see how much of the group a finite ansatz actually covers, we compile a target that acts as a random $SU(5)$ on the $S_{\mathrm{tot}} = 0$ block and as the identity on the others, and maximize $F = |\mathrm{Tr}(V^\dagger U(\boldsymbol{\theta}))|/20$ over $\boldsymbol{\theta}$. With unconstrained pulse areas the best fidelity is $0.80$ at $L = 6$ (30 parameters), $0.93$ at $L = 13$ (65) and $0.998$ at $L = 26$ (130); with the bound $|\theta| \le 0.3$ used in the trained model the corresponding values are $0.78$, $0.84$ and $0.96$.\footnote{Expressivity is monotone in $L$, since a deeper ansatz contains a shallower one by setting the extra angles to zero. At $L = 40$ and $L = 60$ our optimization returns slightly lower fidelities ($0.99$) than at $L = 26$ under the same restart and step budget; this reflects the difficulty of the optimization at larger depth, not a loss of expressivity, and we quote the depths up to $L = 26$ for that reason.} The ansatz thus becomes expressive on this space essentially once $5L$ exceeds the algebra dimension 129, and the depth $L = 6$ used throughout this paper is far from that point. We report this as a property of the ansatz rather than as a requirement for the task: the classification studied here evidently does not need arbitrary transformations of the sector.

\subsection{Data Encoding and Time-Encoded Exchange (TE-EX) Layer}
The transition from classical image data to the quantum Hilbert space is achieved through a hybrid encoding strategy. First, high-dimensional MNIST digits are projected onto a $d=12$ dimensional feature vector $\mathbf{x}$ using Principal Component Analysis (PCA), which retains the leading geometric variance of the dataset.

\begin{figure}[htbp]
    \centering
    \includegraphics[width=1.0\columnwidth]{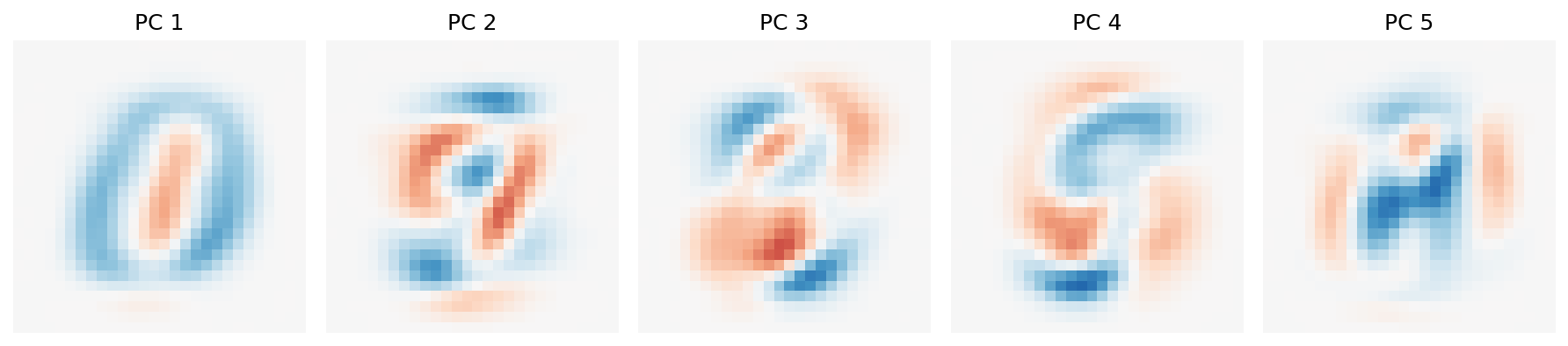} 
    \caption{Top 5 PCA components (eigen-digits) extracted from the MNIST dataset. These components represent the primary geometric structures—such as loops, strokes, and orientations—that are mapped onto the quantum exchange parameters $\mathbf{x}$.}
    \label{fig:pca_input}
\end{figure}

These classical features are subsequently integrated into the quantum evolution by modulating the dimensionless pulse area $\theta_i^{(l)}$ of each exchange link $i$ in each layer $l \in \{1, \dots, 6\}$. The pulse area is proportional to the time-integrated exchange coupling, $\theta_i^{(l)} = \frac{1}{4\hbar}\int J_i(t)\,\mathrm{d}t$ in the units of Eq.~\eqref{eq:generator}, and is set electrically through the amplitude or duration of the barrier-gate voltage pulse; it is the natural control variable of the protocol, and all trainable parameters act on it. 

The mapping between the classical input and the quantum control parameters is defined by a non-linear variational transformation:
\begin{equation}
    \theta_{i}^{(l)}(\mathbf{x}) = \gamma \cdot \tanh \left( \sum_{j=1}^{d} W_{i,l,j} x_j + b_{i,l} \right),
\end{equation}
where $\{W, b\}$ represent the trainable weights and biases of the classical interface. The $\tanh$ activation keeps the pulse areas bounded, $|\theta_i^{(l)}| \le \gamma$. The scale $\gamma = 0.3$ was chosen empirically in preliminary runs, before the systematic search of Sec.~\ref{sec:valsel}; that search shows that the accuracy is not very sensitive to $\gamma$ in the range $0.1$--$0.6$, and that $\gamma = 0.3$ and $\gamma = 0.6$ perform similarly. In units of $\mathbf{S}_i \cdot \mathbf{S}_{i+1}$ the maximal pulse area at $\gamma = 0.3$ is $4\gamma = 1.2$.

The core computational unit of our model is the Time-Encoded Exchange (TE-EX) protocol. In this framework, the system undergoes evolution through $L=6$ sequential layers, each characterized by a series of non-commuting exchange pulses. Within each layer $l$, the total unitary evolution is constructed as an ordered product of local operations:
\begin{equation}
    U_l(\mathbf{x}) = e^{-i h_{N-1,N} \theta_{N-1}^{(l)}(\mathbf{x})} \dots e^{-i h_{1,2} \theta_{1}^{(l)}(\mathbf{x})},
\end{equation}
where $h_{i,i+1}$ is the isotropic link generator of Eq.~\eqref{eq:generator}. 

This analog evolution scheme departs from traditional gate-based quantum circuits. Instead of decomposing a desired transformation into a discrete set of universal gates (e.g., CNOTs), we directly harness the intrinsic, continuous-time dynamics of the semiconductor spin chain. By stacking six layers of these sequential pulses, the non-commutativity ($[H_{i,i+1}, H_{i+1,i+2}] \neq 0$) generates correlations between the pairs. This state mixing provides the input-dependent dynamics on which the classification relies, replacing the external magnetic field gradient by purely electrical, time-domain control.

\subsection{Quantum Feature Extraction and Hybrid Classical Head}
Following the unitary evolution through $L=6$ layers, the final quantum state $|\psi(\mathbf{x}, \boldsymbol{\theta})\rangle$ carries input-dependent correlations between the spins. To translate this quantum representation back into classical information, we perform a series of measurements to extract an 11-dimensional quantum feature vector $\mathbf{h} \in \mathbb{R}^{11}$. This vector is composed of two distinct classes of physical observables:

\begin{enumerate}
    \item \textbf{Local Magnetization (6 features):} We measure the expectation value of the longitudinal spin component at each site, $m_i = \langle S_{z,i} \rangle$ for $i \in \{1, \dots, 6\}$. These values capture the spatial distribution of spin excitations and the local magnetic environment along the chain.
    \item \textbf{Exchange Correlations (5 features):} We measure the nearest-neighbor Heisenberg correlations, $c_{i,i+1} = \langle h_{i,i+1} \rangle = 4\langle \mathbf{S}_i \cdot \mathbf{S}_{i+1} \rangle$. These observables quantify the singlet-triplet character of adjacent dots, which is directly modulated by the TE-EX pulse sequences.
\end{enumerate}

Both observable classes map onto the standard readout primitives of the platform. Each $\langle S_{z,i} \rangle$ is obtained by energy-selective spin-to-charge conversion \cite{Elzerman2004}. Each exchange correlator reduces to a singlet-probability measurement: for a spin pair, $\mathbf{S}_i \cdot \mathbf{S}_j = \tfrac{1}{4}\mathbb{1} - P_S$ with $P_S$ the singlet projector, so $c_{i,j} = 1 - 4\langle P_S \rangle$ is estimated directly by Pauli-spin-blockade readout of the pair \cite{Petta2005}. Because the six $S_z$ observables mutually commute, and the five link correlators split into two sets of disjoint pairs, three measurement settings cover all eleven features. The readout is destructive, so expectation values are accumulated over repeated preparations; the finite-shot analysis of Sec.~\ref{sec:robustness} quantifies exactly this cost, using the exact single-shot eigenvalue spread of each observable.

The extracted features $\mathbf{h}$ are then processed by a classical ``head.'' The expectation values are physically bounded (e.g., $m_i \in [-0.5, 0.5]$) and have different scales, so we place a \textbf{Batch Normalization (BN)} layer immediately after the quantum measurement. The BN layer rescales and centers the features; its effect on accuracy is modest (Sec.~\ref{sec:features}).

The final decision-making process is governed by a multi-layer perceptron (MLP) defined as:
\begin{equation}
    \mathbf{\hat{y}} = \text{Softmax}(\mathbf{W}_2 \cdot \sigma(\text{BN}(\mathbf{W}_1 \mathbf{h} + \mathbf{b}_1)) + \mathbf{b}_2),
\end{equation}
where $\mathbf{W}_1$ maps the 11 quantum features to a 64-dimensional hidden space, and $\sigma$ denotes the ReLU activation function. The entire pipeline—comprising both the quantum pulse parameters and the classical MLP weights—is trained end-to-end for 1000 steps using the Adam optimizer (initial learning rate $3\times 10^{-3}$, cosine-annealing schedule) with a batch size of 128. The encoder and the spin chain together act as a non-linear feature map, and the classical head decodes the eleven observables. All reported accuracies are evaluated on the full 10{,}000-image MNIST test set, which is not used during training. The configuration above (PCA-12, six layers, $\gamma = 0.3$) was fixed in preliminary experiments without a separate validation split; Sec.~\ref{sec:valsel} reports a systematic hyperparameter search on a validation split held out from the training data, which shows that this configuration is not tuned to the test set and identifies a larger configuration with higher accuracy.

\section{Results}

\subsection{Training Convergence and Classification Accuracy}
\begin{figure}[htbp]
    \centering
    \includegraphics[width=1.0\columnwidth]{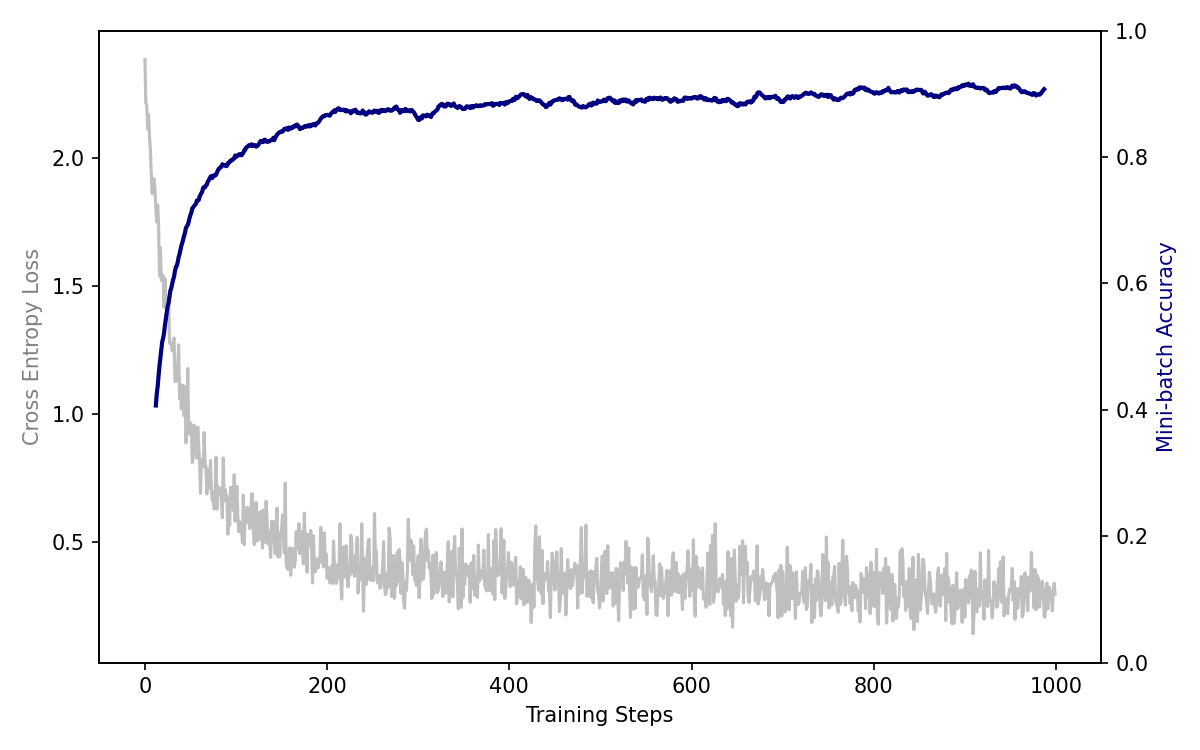}
    \caption{Training loss (grey) and smoothed mini-batch accuracy (navy) over 1000 steps. Batch size 128 with a cosine-annealing schedule; the final full-test-set accuracy is 90.89\%.}
    \label{fig:loss_curve}
\end{figure}

The hybrid TE-EX model was trained for 1000 steps with a batch size of 128. As shown in Fig.~\ref{fig:loss_curve}, training converged smoothly. In the initial phase (0--200 steps), the cross-entropy loss decreased sharply from 2.4 to below 0.7, reflecting the rapid adaptation of the classical-quantum interface to the PCA-reduced features, and settled near 0.2--0.3 by the end of training.

\begin{figure}[htbp]
    \centering
    \includegraphics[width=1.0\columnwidth]{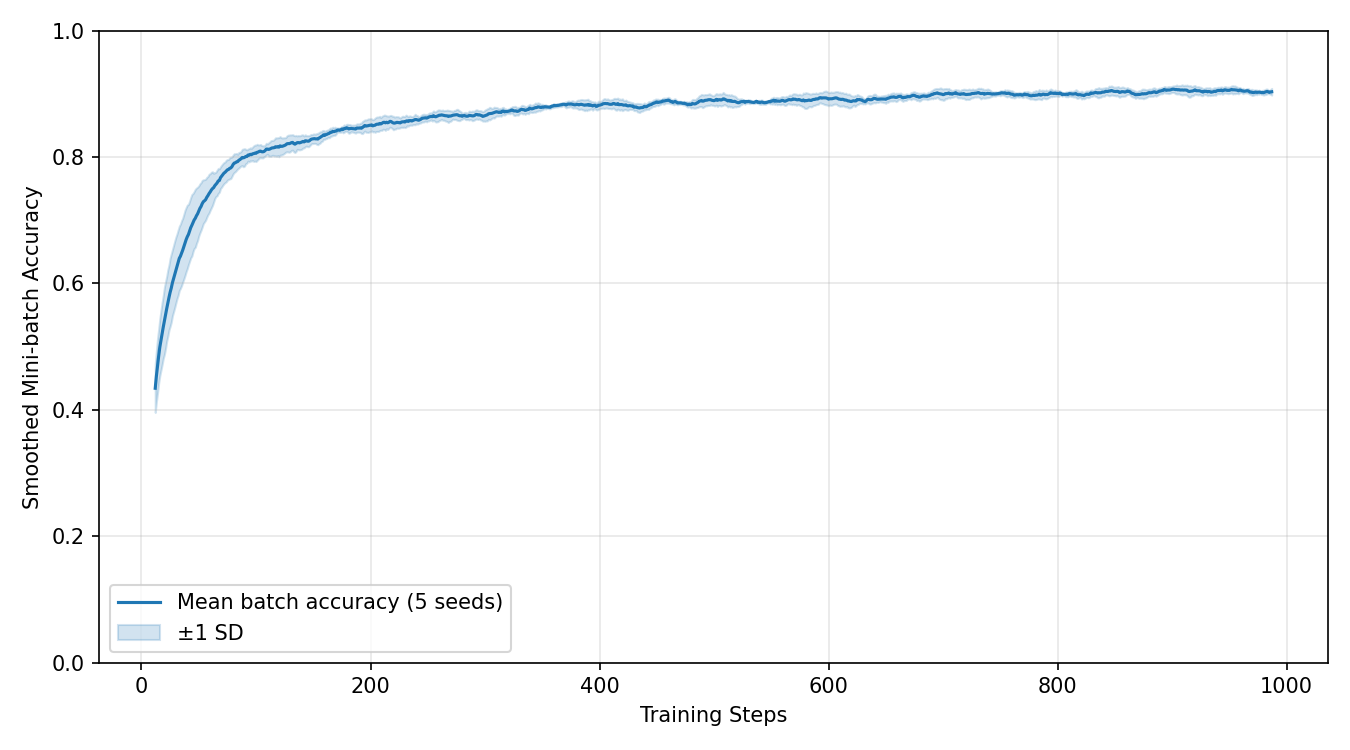}
    \caption{Training stability over five independent random seeds (paper configuration). Shown is the smoothed mini-batch accuracy (mean $\pm$ one standard deviation); the resulting full-test-set accuracy is $90.99\% \pm 0.24\%$.}
    \label{fig:stability}
\end{figure}

The trained model (seed 42) achieves a final accuracy of \textbf{90.89\%} on the full 10{,}000-image MNIST test set. To assess seed dependence, the entire training procedure was repeated with five random seeds (42--46, the first being the run above), yielding a test accuracy of $90.99\% \pm 0.24\%$ (mean $\pm$ one population standard deviation; Fig.~\ref{fig:stability}), so the result is stable across seeds. This accuracy is obtained with 6 spins and without any local magnetic field gradient ($\Delta B_z$): the non-commuting sequential exchange pulses ($[H_{i,i+1}, H_{i+1,i+2}] \neq 0$) provide enough state mixing to map the 12-dimensional ``eigen-digits'' into classifiable states of the $M_z = 0$ sector.

\begin{figure}[htbp]
    \centering
    \includegraphics[width=1.0\columnwidth]{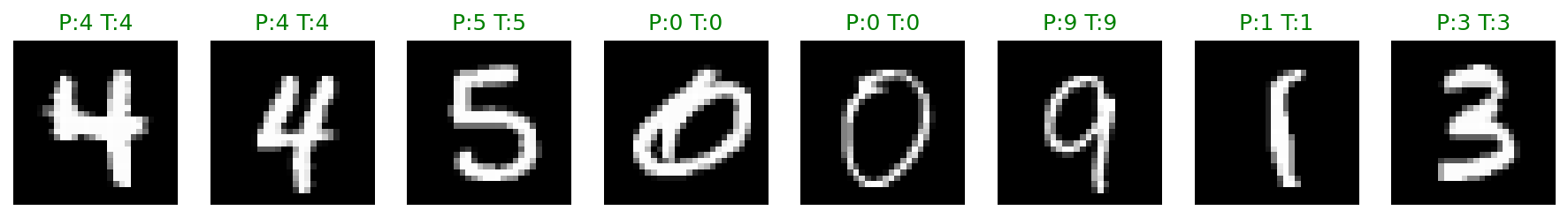}
    \caption{Representative model predictions on the MNIST test set. Correct classifications are indicated in green, misclassifications in red.}
    \label{fig:mnist_examples}
\end{figure}

The model's predictive capability is qualitatively evaluated in Fig.~\ref{fig:mnist_examples}, which displays random samples from the test set alongside their predicted (P) and true (T) labels.

\subsection{Classical Baselines and Ablations}
\label{sec:baselines}
To isolate the contribution of the quantum feature map, we compare the TE-EX model against classical baselines that receive the identical 12-dimensional PCA input, as well as ablated variants of the model itself. All models are trained with the same optimizer, batch size, and step budget, and evaluated on the full test set (Table~\ref{tab:baselines}).

\begin{table}[htbp]
\caption{Test accuracy on the full MNIST test set (10{,}000 images). All learned models use identical PCA-12 inputs, optimizer, and training budget (Adam, batch 128, 1000 steps).}
\label{tab:baselines}
\begin{ruledtabular}
\begin{tabular}{lc}
Model & Test acc. (\%) \\
\colrule
Logistic regression on PCA-12 (linear) & 83.2 \\
Classical MLP head on PCA-12 & 91.3 \\
\colrule
TE-EX, frozen random pulses + MLP head & 53.0 \\
TE-EX, trained pulses + linear readout & 88.1 \\
TE-EX, trained pulses + MLP head (full) & \textbf{90.9} \\
\end{tabular}
\end{ruledtabular}
\end{table}

Three observations follow. First, training the pulse parameters matters: freezing the exchange pulses at random values---i.e., operating the chain as a static quantum reservoir---reduces the accuracy to 53.0\%, below the linear classical baseline. The frozen-pulse baseline uses random weights of scale $0.1$ without any tuning of that scale, so it is a lower bound on what a reservoir with an optimized readout could reach; what it shows is that the accuracy of the TE-EX model comes from the \emph{learned}, input-dependent dynamics rather than from the dynamics alone. Second, under a strictly linear readout the trained model (88.1\%) outperforms the best linear classifier acting directly on the same PCA inputs (83.2\%) by about five percentage points, even though the features compress the 12-dimensional input into 11 observables. The feature map is therefore non-linear---but the non-linearity is not supplied by the exchange dynamics alone: every pulse area is $\theta_{l,i} = \gamma \tanh(b_{l,i} + \mathbf{w}_{l,i}\cdot\mathbf{x})$, so the encoder already contains 30 $\tanh$ units before the chain is reached. The combination of the non-linear $\tanh$ parameterization and the learned, input-dependent exchange dynamics implements a trainable non-linear feature map; Section~\ref{sec:contribution} separates the two contributions. Third, the full hybrid model (90.9\%) is comparable to a classical MLP of the same head architecture applied directly to the PCA features (91.3\%). The aim of this work is not to claim an accuracy advantage over classical networks, but to show that a minimal, gradient-free spin-chain substrate---with the classical post-processing held fixed---reaches the same performance class, which is the relevant benchmark for hardware-efficient QML on near-term spin-qubit devices.

\subsection{Hyperparameter Selection on a Validation Split}
\label{sec:valsel}
The configuration used above (PCA-12, $L = 6$, $\gamma = 0.3$) was fixed tentatively in preliminary experiments without a separate validation split. To check that it is not tuned to the test set, and to see how the three main hyperparameters affect the accuracy, we ran a post-hoc grid search with a proper protocol: the 60{,}000 training images were split once into 50{,}000 for training and 10{,}000 for validation, PCA was fit on the 50{,}000 only, and every combination of PCA dimension $d \in \{8, 12, 16\}$, depth $L \in \{4, 6, 8\}$ and pulse scale $\gamma \in \{0.1, 0.3, 0.6\}$ was trained on the 50{,}000 with all other settings as above and scored on the validation split (Table~\ref{tab:valsel}). The configuration with the best validation accuracy was then retrained on the full 60{,}000 training images and evaluated once on the test set.

\begin{table}[htbp]
\caption{Validation accuracy (\%) of the 27 configurations of the grid search (trained on 50{,}000 images, evaluated on 10{,}000 validation images held out from the training set; seed 42). The configuration used in the rest of the paper is marked with an asterisk; the best configuration on the validation split is in bold.}
\label{tab:valsel}
\begin{ruledtabular}
\begin{tabular}{llccc}
PCA dim. & Layers & $\gamma = 0.1$ & $\gamma = 0.3$ & $\gamma = 0.6$ \\
\colrule
8  & 4 & 86.2 & 86.1 & 85.2 \\
8  & 6 & 85.9 & 86.0 & 86.7 \\
8  & 8 & 85.8 & 86.7 & 86.8 \\
\colrule
12 & 4 & 89.6 & 89.8 & 89.7 \\
12 & 6 & 90.0 & 90.2$^{*}$ & 90.3 \\
12 & 8 & 89.7 & 90.8 & 90.7 \\
\colrule
16 & 4 & 90.7 & 91.6 & 91.3 \\
16 & 6 & 91.0 & 92.4 & 92.4 \\
16 & 8 & 90.8 & 91.3 & \textbf{92.7} \\
\end{tabular}
\end{ruledtabular}
\end{table}

The paper configuration ranks 13th of 27 on the validation split (90.2\%), in the middle of the grid, so it was not selected by looking at the test set. The dominant factor is the PCA dimension: the best PCA-8, PCA-12 and PCA-16 configurations reach 86.8\%, 90.8\% and 92.7\%, and every PCA-12 configuration beats every PCA-8 configuration, i.e., the accuracy is limited mainly by the classical input information passed to the chain. Depth and pulse scale have smaller effects. Neither $\gamma = 0.3$ nor $\gamma = 0.6$ is consistently better---their difference is within $0.3$ points in six of nine cases---while $\gamma = 0.1$ is the worst of the three in eight of nine cases; the smaller pulse scale used in the rest of the paper is therefore not needed for trainability at this system size. The validation-selected configuration (PCA-16, $L = 8$, $\gamma = 0.6$), retrained on the full training set, reaches \textbf{93.25\%} on the test set. We keep PCA-12, $L = 6$, $\gamma = 0.3$ as the reference configuration for all other analyses in this paper, since the baselines, ablations and noise studies were run with it; the validation-selected result shows that the numbers reported here are conservative rather than optimistic.

\subsection{Which Observables Are Needed?}
\label{sec:readout}
The eleven features are read out by two different primitives. The six local magnetizations $\langle S_{z,i} \rangle$ require energy-selective spin-to-charge conversion \cite{Elzerman2004}, whereas the five correlators $c_{i,i+1} = 1 - 4\langle P_S \rangle$ are singlet-probability measurements of a pair, obtained by Pauli spin blockade \cite{Petta2005}---the standard readout of singlet-triplet devices. Since the two carry different experimental cost, it is worth asking whether both are necessary.

Retraining the paper configuration with the classical head restricted to one group at a time (five seeds each, everything else unchanged) gives $89.4\% \pm 0.5\%$ from the five pair correlators alone and $87.1\% \pm 0.6\%$ from the six magnetizations alone, against $90.5\% \pm 0.2\%$ when all eleven are used under the identical protocol.\footnote{The eleven-feature control is quoted at $90.5\%$ rather than the $90.89\%$ of Table~\ref{tab:baselines} because the head is constructed after the model in this experiment, which changes the order in which random numbers are drawn; the three arms are directly comparable with one another.} The singlet-probability channel alone therefore costs about one percentage point, while carrying fewer features than the magnetizations it outperforms. The protocol can consequently be operated with Pauli spin blockade as its only readout channel, which removes the need for single-spin readout in an experimental realization.

\subsection{What Does the Spin Chain Contribute?}
\label{sec:contribution}
The encoder of the TE-EX model is already a classical non-linear map: the $L(N-1) = 30$ pulse areas are $\gamma \tanh(b + \mathbf{w}\cdot\mathbf{x})$, parameterized by $360 + 30 = 390$ numbers. The chain maps those 30 angles onto 11 observables, and a BatchNorm plus a linear layer reads them out (142 parameters; 532 in total for the linear-readout model of Table~\ref{tab:baselines}). Two families of controls, trained with the identical optimizer and budget, separate the contribution of the encoder from that of the chain.

First, purely classical $\tanh$ networks of varying width, with the same optimizer and budget. At a matched parameter count (510 against 532) such a network reaches $86.5\% \pm 0.3\%$, below the $88.1\%$ of the TE-EX model with a linear readout; given the same 30 $\tanh$ units but a wider readout (760 parameters) it reaches $88.4\% \pm 0.3\%$. Whether the chain appears to help therefore depends on whether one matches the parameter count or the structure of the encoder.

Second, and more directly, we keep the encoder and the readout exactly as they are and replace only the chain by a fixed, untrained map from the 30 angles to 11 features, so that every arm has 532 trainable parameters and the arms differ solely in what performs that step. A fixed random linear map gives $88.5\% \pm 0.5\%$; a fixed random quadratic form $f_k = \boldsymbol{\theta}^{\mathsf{T}} M_k \boldsymbol{\theta}$---a deliberately simple non-linear map with no dynamics behind it; it is quadratic in the pulse areas themselves, whereas the observables of the chain are quadratic in $\psi(\boldsymbol{\theta})$ and hence trigonometric in $\boldsymbol{\theta}$---gives $90.2\% \pm 0.3\%$ (five independent random maps, three seeds each), against $88.1\%$ for the chain itself at the reference seed and $89.0\% \pm 0.6\%$ over the same five seeds.

We therefore do not claim that the exchange dynamics constitute a better feature map than a simple classical non-linear map; on this task they do not. The $\tanh$ encoder followed by a fixed non-linear compression to 11 features already lifts the model above the linear baseline, and the exchange dynamics are not needed for that step. The random quadratic form is not a handicapped stand-in for the chain: its $11 \times 30 \times 30$ fixed coefficients make its fixed layer less constrained than the chain, whose fixed part is the Hamiltonian structure. What the chain provides is that its feature map is executed by the native, all-electrical dynamics of the hardware, under constraints---unitarity, conservation of $M_z$ and $\mathbf{S}_{\mathrm{tot}}$, bounded observables, and the limited depth quantified in Sec.~\ref{sec:reachable}---that an unconstrained random quadratic form does not have to respect. Those constraints are the price of physical realizability, and they plausibly account for the gap. We regard this as the honest reading of the ablations: the contribution of the protocol is hardware efficiency, not an advantage of the quantum feature map over its classical counterpart.

\subsection{Robustness to Control Noise and Finite Sampling}
\label{sec:robustness}
Two imperfections dominate any experimental realization of the protocol: fluctuations of the exchange coupling caused by charge noise, and the statistical error of estimating expectation values from a finite number of measurement shots. We evaluate both at test time on the trained model of Table~\ref{tab:baselines}, using the full 10{,}000-image test set and five independent noise realizations per setting; the model itself is trained noiselessly.

\begin{figure}[htbp]
    \centering
    \includegraphics[width=1.0\columnwidth]{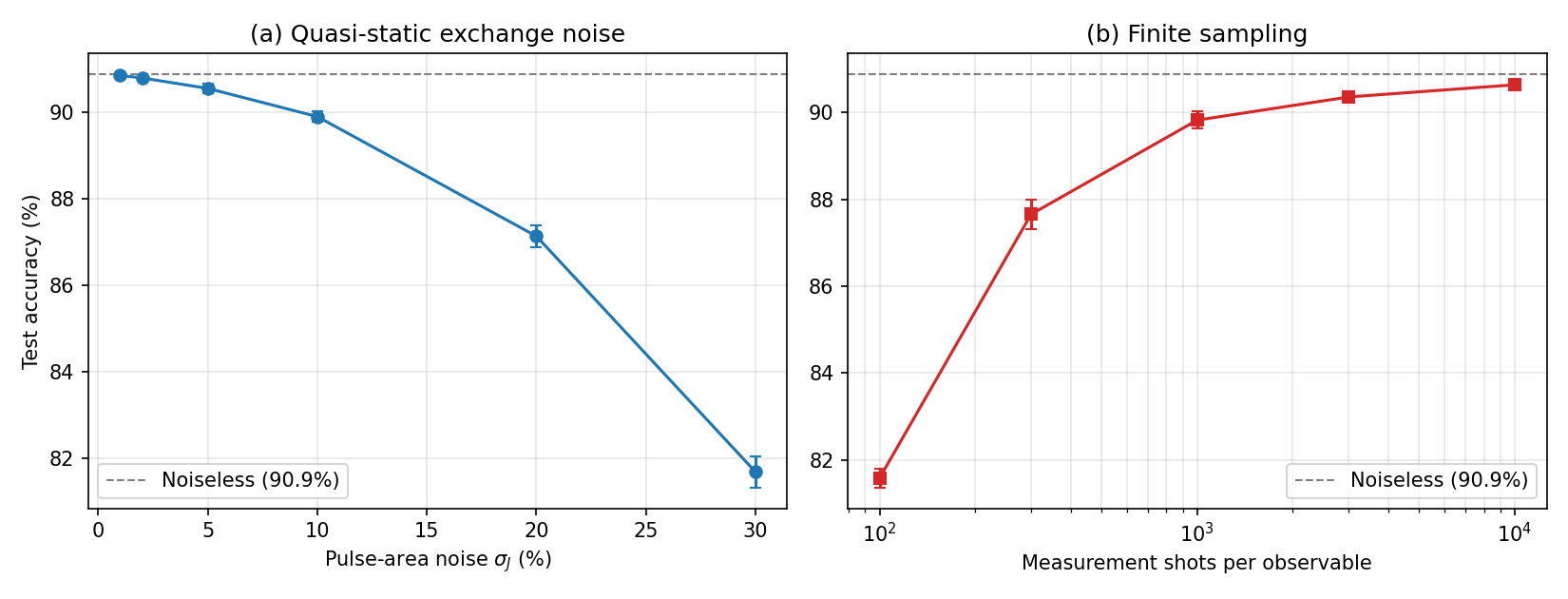}
    \caption{Robustness of the trained TE-EX model (full test set; mean $\pm$ one standard deviation over five noise realizations). (a) Test accuracy under quasi-static multiplicative pulse-area noise of relative strength $\sigma_J$. (b) Test accuracy when every observable is estimated from a finite number of measurement shots. Dashed lines mark the noiseless accuracy (90.9\%).}
    \label{fig:robustness}
\end{figure}

\textit{Quasi-static exchange noise.}---Because the exchange coupling depends exponentially on the barrier-gate voltage, charge noise translates predominantly into \emph{multiplicative} fluctuations of the pulse area \cite{Petta2005, Burkard2023}. We therefore replace every pulse area by $\theta \to \theta(1+\varepsilon)$ with $\varepsilon \sim \mathcal{N}(0, \sigma_J^2)$ drawn independently for every link, layer, and input sample, modelling shot-to-shot quasi-static noise. As shown in Fig.~\ref{fig:robustness}(a), the accuracy degrades gradually: the loss is below $0.1$ percentage points for $\sigma_J \le 2\%$, and the model still reaches $90.6\%$ at $\sigma_J = 5\%$, $89.9\%$ at $10\%$, and $87.1\%$ at $20\%$. Even at $\sigma_J = 30\%$ the accuracy ($81.7\%$) remains close to that of the best linear classifier operating on \emph{noiseless} classical inputs ($83.2\%$), indicating that the learned feature map does not rely on finely tuned interference.

\textit{Finite measurement shots.}---In experiment, each of the 11 observables is estimated from $n$ repeated preparations. We model this by adding Gaussian noise with the exact single-shot quantum variance, $\hat{h}_k = h_k + \mathcal{N}\big(0, (\langle O_k^2\rangle - \langle O_k \rangle^2)/n\big)$, the central-limit approximation of projective sampling. Figure~\ref{fig:robustness}(b) shows that $10^3$ shots per observable retain $89.8\%$ accuracy, while $10^4$ shots ($90.6\%$) are nearly indistinguishable from the exact expectation values. Since the six $\langle S_{z,i} \rangle$ mutually commute and the five link correlators split into two sets of disjoint---hence commuting---links, all 11 observables can be covered by three measurement settings. Combining both imperfections at a representative operating point ($\sigma_J = 5\%$ and $10^3$ shots) yields $89.8\% \pm 0.1\%$.

This analysis captures control and sampling errors but not decoherence during the pulse sequence itself; as estimated in Sec.~\ref{sec:hardware}, the total evolution time is short enough that such effects are expected to be subdominant, and a microscopic treatment is left for future work.

\subsection{Quantum Feature Analysis}
\label{sec:features}
\begin{figure}[htbp]
    \centering
    \includegraphics[width=1.0\columnwidth]{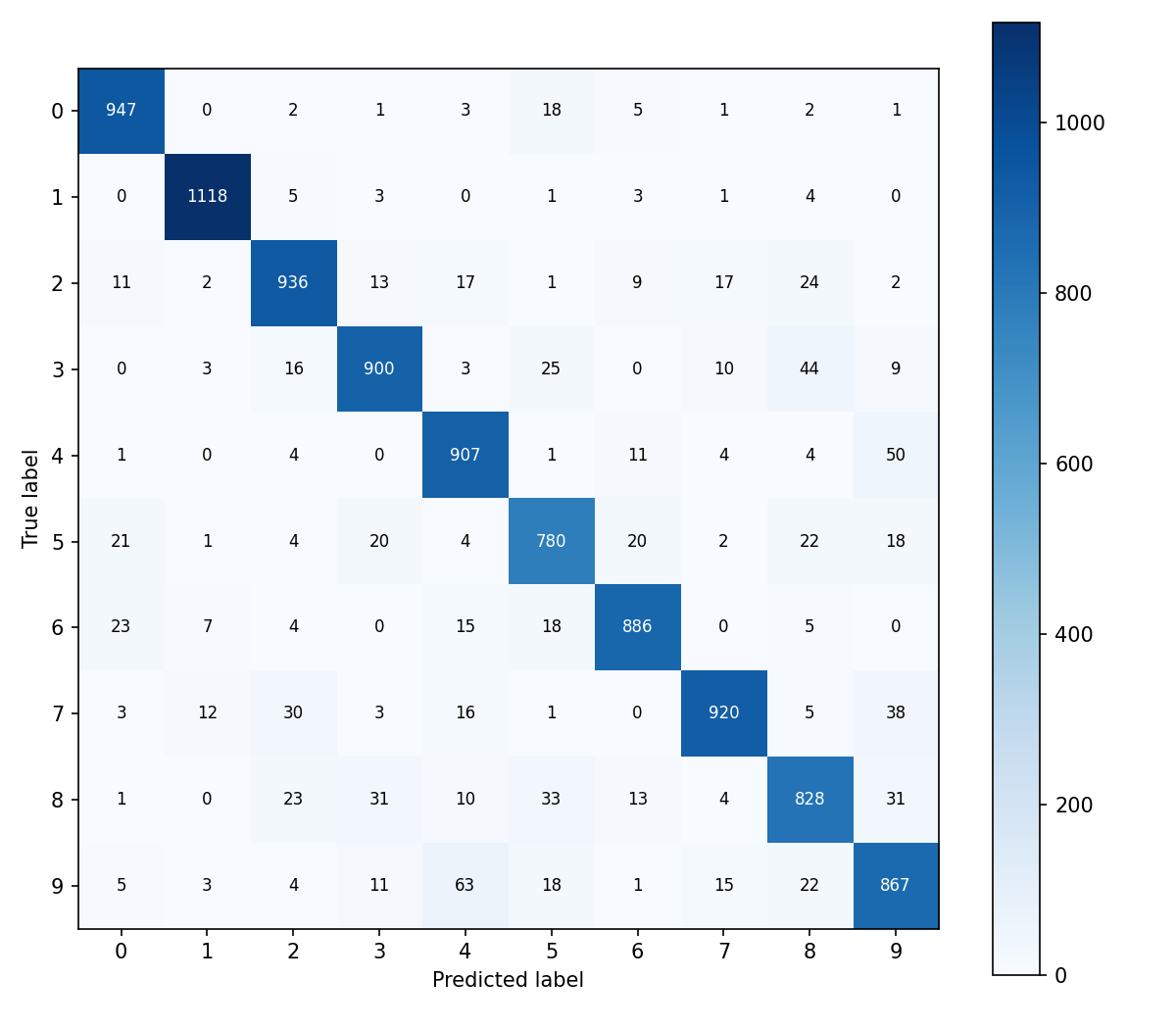}
    \caption{Confusion matrix of the 6-spin TE-EX model on the full MNIST test set (10{,}000 images, overall accuracy 90.89\%). Per-class recall ranges from 85.0\% (``8'') to 98.5\% (``1''); the largest confusions are $9 \to 4$, $4 \to 9$ and $3 \to 8$.}
    \label{fig:confusion_matrix}
\end{figure}

The confusion matrix of the reference model (Fig.~\ref{fig:confusion_matrix}) shows per-class recall between 85.0\% (``8'') and 98.5\% (``1''); ``0'' (96.6\%), ``6'' (92.5\%) and ``4'' (92.4\%) are also above average, while ``8,'' ``9'' (85.9\%) and ``5'' (87.4\%) are the hardest classes. The largest confusions are between visually similar digits: $9 \to 4$ (63 images), $4 \to 9$ (50), $3 \to 8$ (44) and $7 \to 9$ (38).

The Batch Normalization (BN) layer has a small effect. Since the quantum expectation values are bounded within a narrow range (e.g., $|\langle S_z \rangle| \le 0.5$), the BN layer rescales them to a distribution suited to the ReLU activation in the classical head. Removing the BN layer, with everything else unchanged (single seed), reduces the final test accuracy from 90.9\% to 89.1\%.

\subsection{Spin Dynamics and Trajectory Expressibility}
The internal dynamics of the 6-spin chain were analyzed to understand the ``feature-to-state'' mapping. Starting from the $|010101\rangle$ antiferromagnetic initial state, the 6-layer sequence of 30 interleaved exchange pulses drives the many-body wavefunction through a data-dependent trajectory in the 20-dimensional $M_z = 0$ subspace.

\begin{figure}[t]
    \centering
    \includegraphics[width=1.0\columnwidth]{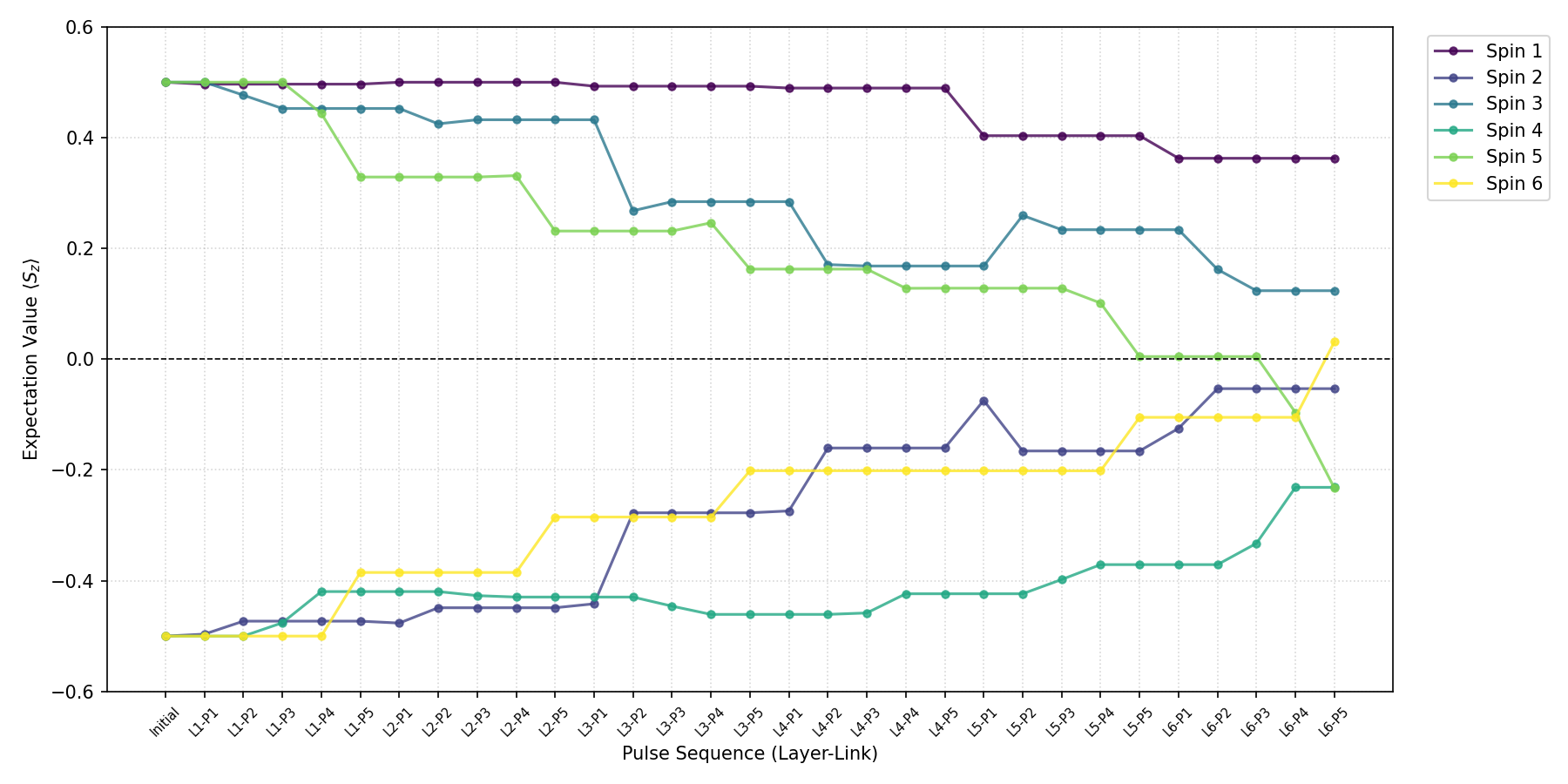}
    \caption{Evolution of the local magnetization $\langle S_{z,i} \rangle$ of each spin during the 30-pulse TE-EX sequence for a representative test sample. Starting from the N\'eel state $|010101\rangle$, the interleaved non-commuting exchange pulses progressively mix the spin populations, driving the state through a data-dependent trajectory in the $M_z=0$ subspace.}
    \label{fig:spin_evolution}
\end{figure}

Because adjacent exchange terms do not commute, the pulses do not compose into independent pair rotations, and differences in the input PCA components are turned into different final states. Figure~\ref{fig:spin_evolution} shows one such trajectory: the local magnetizations, which start at $\pm 1/2$, are gradually mixed by the 30 pulses.

\subsection{Learned Weights}
\begin{figure}[htbp]
    \centering
    \includegraphics[width=1.0\columnwidth]{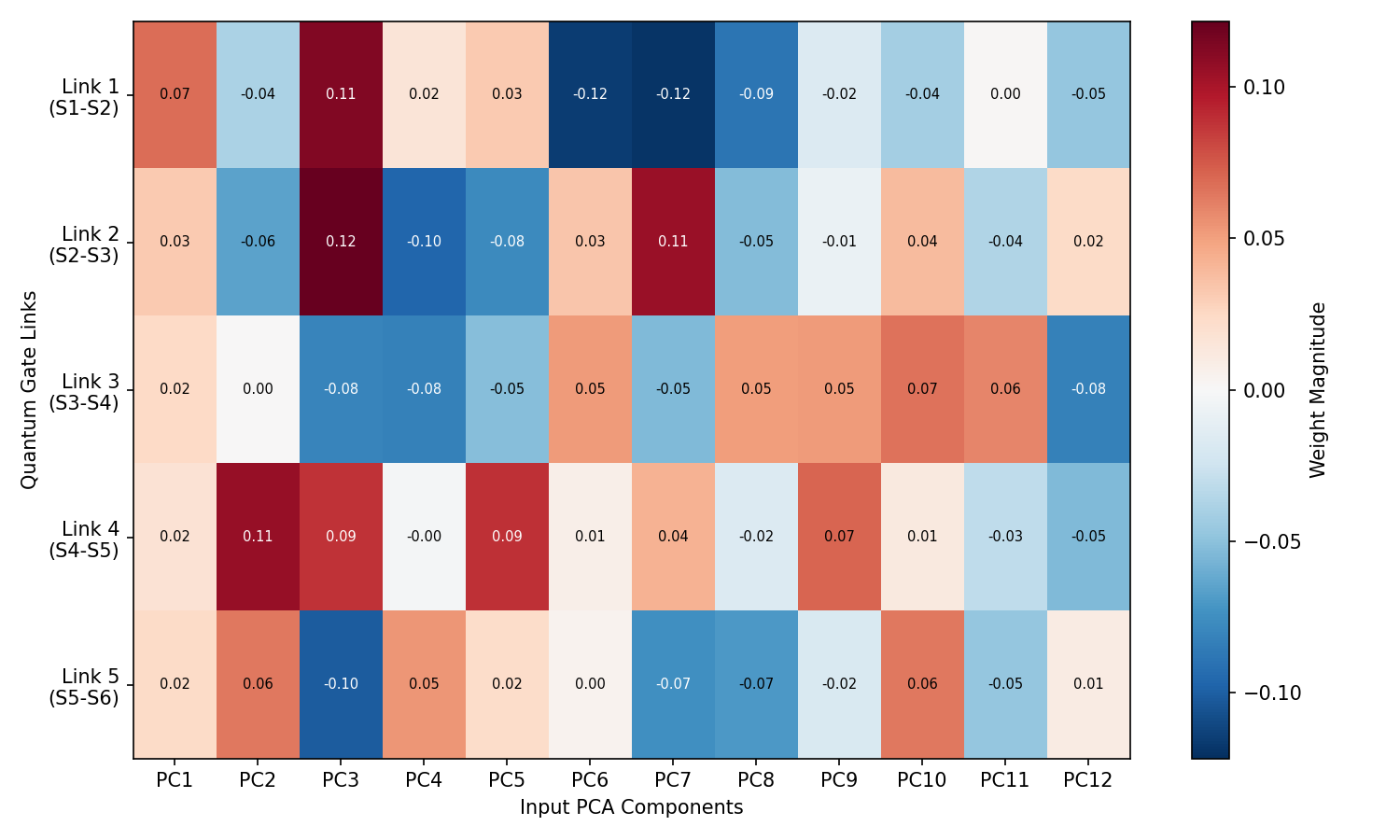}
    \caption{Heatmap of the learned weights $\mathbf{W}$ in the first variational layer. All weights are small ($|W| \le 0.19$ in every layer) and no PCA component couples to a single link only; see text.}
    \label{fig:weight_heatmap}
\end{figure}

Figure~\ref{fig:weight_heatmap} shows the learned weight matrix $\mathbf{W}$ of the first layer, which maps the 12 PCA components to the pulse areas of the five links.

The learned weights are small and spread out rather than sparse: $|W| \le 0.19$ in all six layers, and in the first layer the mean $|W|$ over the five links ranges from $0.03$ (PC1) to $0.10$ (PC3), with PC3 coupling almost uniformly to all links. Averaged over the six layers, no component dominates (mean $|W|$ between $0.03$ and $0.05$ for every component). Each pulse area is therefore a small, weighted combination of many PCA components rather than a function of one or two of them, and the first-layer heatmap alone does not reveal a simple correspondence between individual eigen-digits and individual links. We do not attempt a more detailed interpretation.

\subsection{Hardware Efficiency and Comparison}
\label{sec:hardware}
Unlike traditional three-spin exchange-only (EO) or gradient-dependent singlet-triplet (S-T) qubits, our TE-EX model does not require maintaining a rigid logical encoding or complex magnetic infrastructure. Because the classical head reads raw physical observables, the model uses the whole $M_z = 0$ sector and there is no ``leakage'' subspace to avoid.

The analog nature of the TE-EX protocol keeps the control sequence short: the whole feature map consists of 30 exchange pulses, each of which is a native operation of the platform, whereas a gate-based implementation would decompose the same unitary into a sequence of discrete two-qubit gates with a correspondingly longer total duration. The required pulses are also experimentally modest: with pulse areas bounded by $|\theta| \le \gamma_{\mathrm{eff}} = 1.2$ and exchange couplings electrically tunable from the sub-MHz to the GHz range \cite{Petta2005, Burkard2023}, a single pulse corresponds to a duration of order $\theta \hbar / J \approx 2\,\mathrm{ns}$ at $J/h = 100\,\mathrm{MHz}$, so the full 30-pulse sequence completes within roughly $10^2\,\mathrm{ns}$---orders of magnitude below the coherence times of electron spins in isotopically purified silicon \cite{Burkard2023}. Together with the noise tolerance of Sec.~\ref{sec:robustness}, this suggests that the protocol is compatible with current silicon quantum-dot technology, although an experimental demonstration is beyond the scope of this work.

\subsection{Visualization of the Quantum Feature Space}
To visualize the feature space, we applied \textbf{t-SNE (t-distributed Stochastic Neighbor Embedding)} to the 11-dimensional feature vectors $\mathbf{h}$ extracted from the test set.

\begin{figure}    \centering
    \includegraphics[width=1.0\columnwidth]{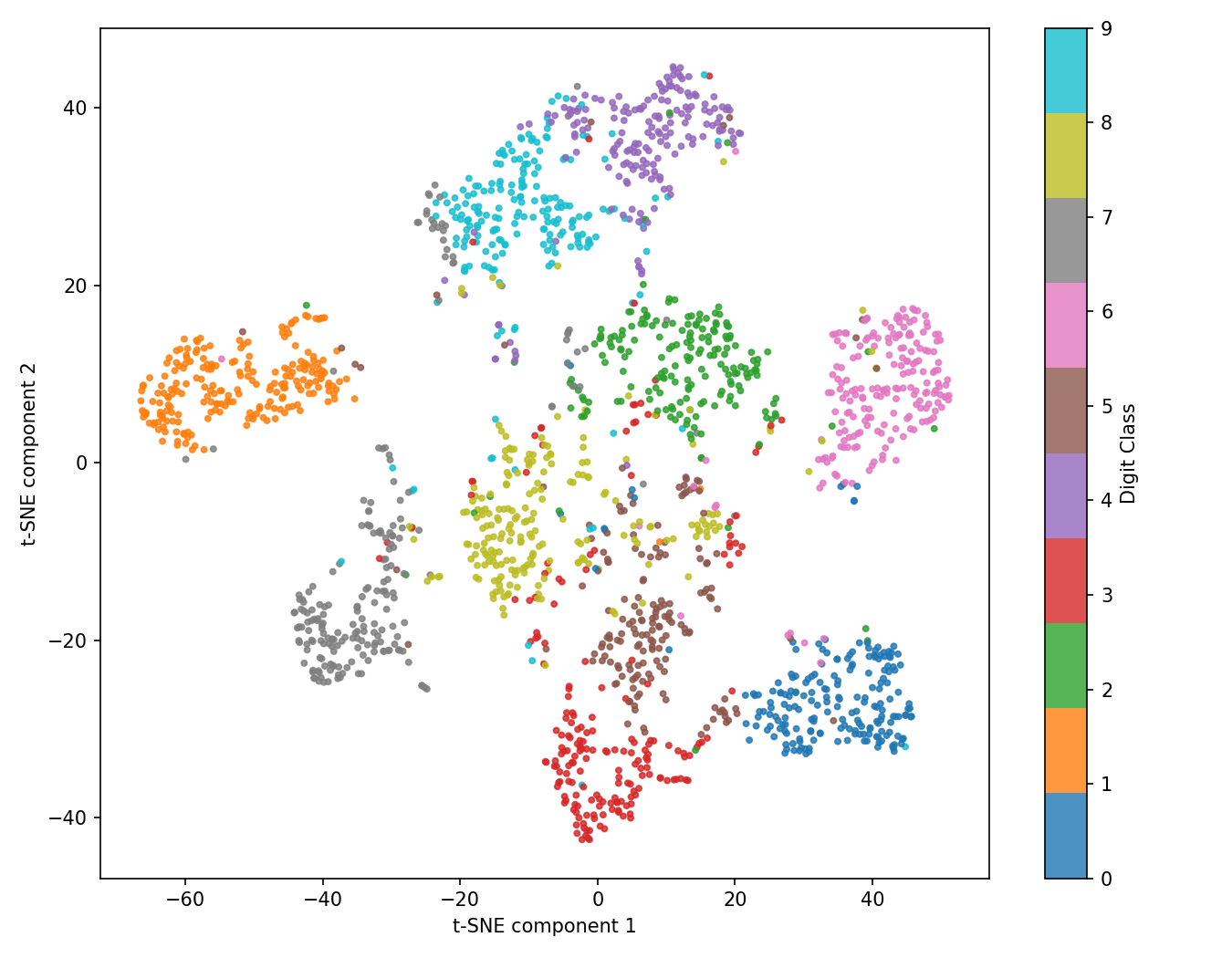} 
    \caption{t-SNE visualization of the 11-dimensional quantum feature vectors of the test set, colored by digit class.}
    \label{fig:tsne}
\end{figure}

As shown in Fig.~\ref{fig:tsne}, the t-SNE projection shows class-specific clusters: the exchange-only dynamics map the ``eigen-digits'' into largely separated regions of the feature space. The proximity of certain clusters (e.g., ``4'' and ``9'') corresponds to the misclassifications seen in the confusion matrix, i.e., the feature map preserves the similarity structure of the input data.

\subsection{Quantum Signatures of Digit Classes}
\begin{figure*}[t]    \centering
    \includegraphics[width=0.9\textwidth]{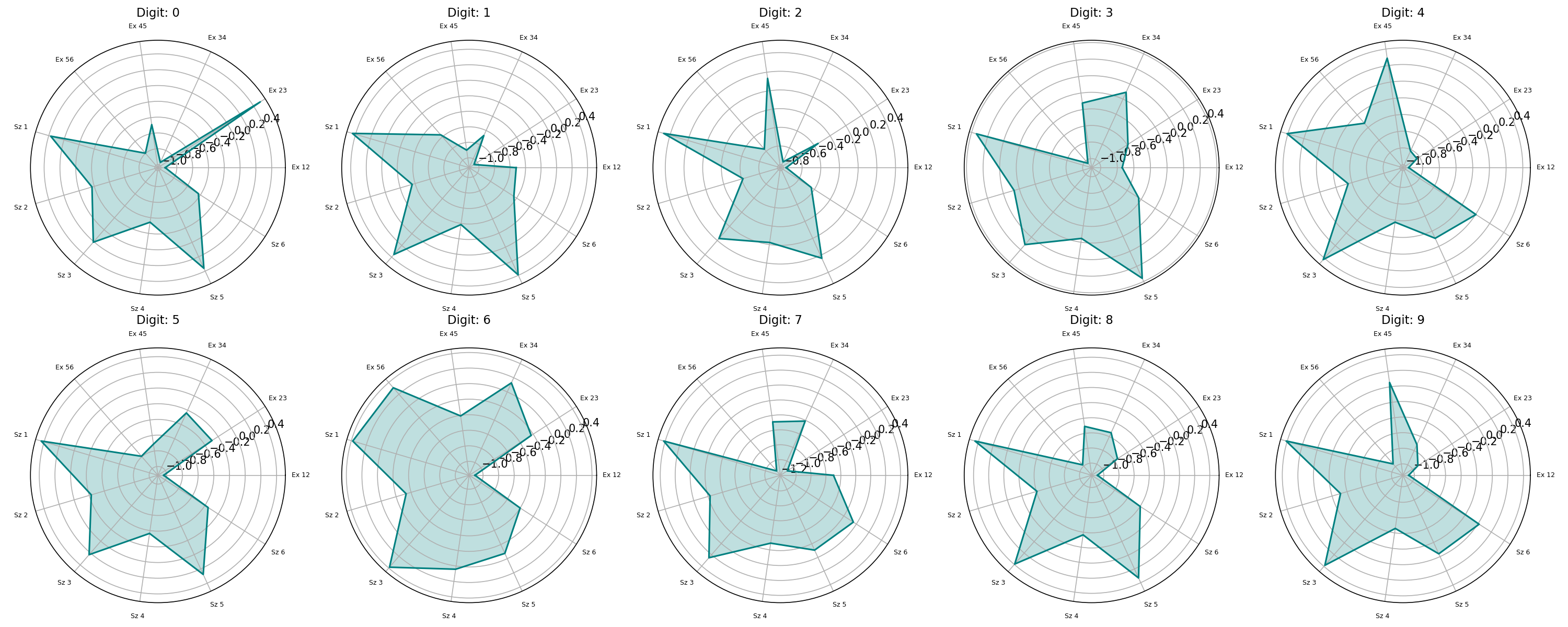}    \caption{\textbf{Class-averaged quantum feature signatures.} Radar charts of the 11 observables (five link correlators $\langle h_{i,i+1} \rangle$, labelled Ex, and six local magnetizations $\langle S_{z,i} \rangle$, labelled Sz) averaged over all test images of each digit. The classes differ mainly in the link correlators; see text.}
    \label{fig:radar_chart}
    \vspace{10pt}\end{figure*}

Averaging the 11 observables $\langle S_{z,i} \rangle$ and $\langle h_{i,i+1} \rangle$ over all test images of one digit gives a class-averaged ``quantum signature'' (Fig.~\ref{fig:radar_chart}).

The class averages differ mainly in the link correlators. For digit ``1'' the final state stays close to the initial N\'eel state: the magnetization remains staggered ($\langle S_{z,i} \rangle \approx \pm 0.4$) and all five correlators are negative ($-1.05$ to $-0.52$). For digit ``0'' the correlators vary strongly from link to link (from $-0.97$ to $+0.50$) and the staggered magnetization is partly washed out (down to $|\langle S_{z,2} \rangle| = 0.17$). The other digits lie between these cases; ``8'' is the class closest to ``1'' in this representation. These class-dependent averages show that the pulse sequence performs a structured, input-dependent mapping rather than randomizing the state, but they are averages over thousands of images and are not by themselves sufficient for classification; the classical head operates on the per-image feature vectors.

\section{Conclusion}

In this work, we have studied numerically a Quantum Machine Learning (QML) architecture based on a linear chain of six semiconductor spin qubits. For the learning task considered, the ``Time-Encoded Exchange'' (TE-EX) protocol removes two hardware requirements of spin-based quantum computing: local magnetic field gradients ($\Delta B_z$) and leakage-free logical encodings.

The main result is that the non-commutativity of interleaved Heisenberg exchange pulses ($[H_{i,i+1}, H_{i+1,i+2}] \neq 0$), without any magnetic gradient, provides enough state mixing within the $M_z = 0$ sector for MNIST classification. With six variational layers the model reaches $90.99\% \pm 0.24\%$ on the full MNIST test set across five seeds, and a hyperparameter search on a validation split held out from the training data identifies a larger configuration (PCA-16, eight layers, $\gamma = 0.6$) that reaches $93.25\%$. The ablations indicate that the pulse areas have to be learned---freezing them at random values reduces the accuracy to 53.0\%, and with a strictly linear readout the trained model (88.1\%) outperforms the best linear classifier on the identical PCA inputs (83.2\%). They do not, however, support attributing the non-linearity of the feature map to the exchange dynamics: driven by the same encoder and read out identically, a fixed, untrained quadratic map of the pulse areas reaches 90.2\%. What the chain contributes is that this feature map is executed by the native, all-electrical dynamics of the hardware, with a readout that Pauli spin blockade alone can supply (89.4\% from the five pair correlators). The dynamics of the spin chain under all-electrical exchange control can thus substitute for external magnetic infrastructure in this task, reaching the same performance class as an equally sized classical neural network on the same inputs. The protocol also tolerates the imperfections expected in experiment: accuracy remains at $89.9\%$ under $10\%$ quasi-static pulse-area noise and at $89.8\%$ when every observable is estimated from $10^3$ measurement shots.

The analysis of the feature space---t-SNE projections and class-averaged ``quantum signatures''---shows that the 11 observables (local magnetizations and nearest-neighbor correlators) separate the digit classes and preserve the similarity structure of the inputs. The Batch Normalization layer at the quantum-classical interface gives a modest improvement (90.9\% versus 89.1\% without it).

The protocol avoids universal gate decomposition and relies on native spin interactions, so the required coherence time is short (about $10^2\,\mathrm{ns}$ for the 30-pulse sequence at $J/h = 100\,\mathrm{MHz}$). Future work will extend the quasi-static control-noise analysis of Sec.~\ref{sec:robustness} to microscopic decoherence models---finite-bandwidth charge-noise spectra and hyperfine-induced dephasing during the pulse sequence---and will address the scaling of the architecture to larger multi-dot arrays. Our results suggest a hardware-efficient route to quantum machine learning on near-term semiconductor spin-qubit hardware.

\begin{acknowledgments}
The author thanks George Monma for reviewing, correcting and checking the code. Anthropic's Claude (Claude Code) was used to assist with code verification, independent re-checks of the numerical results, and editing of the manuscript; the author takes full responsibility for the content of this paper.
\end{acknowledgments}

\section*{Code Availability}
The simulation and training code that reproduces all results and figures in this paper, together with the raw numerical results, is openly available at \url{https://github.com/minatoyuichiro/teex-mnist}. The validation-split hyperparameter search of Sec.~\ref{sec:valsel} and the Batch-Normalization ablation of Sec.~\ref{sec:features} are reproduced by \texttt{validation\_selection.py} and \texttt{bn\_ablation.py}; the readout ablation of Sec.~\ref{sec:readout}, the classical controls of Sec.~\ref{sec:contribution}, and the Lie-algebra dimension and finite-depth expressivity study of Sec.~\ref{sec:reachable} are reproduced by \texttt{readout\_ablation.py}, \texttt{classical\_controls.py}, \texttt{dla\_dimension.py} and \texttt{ansatz\_expressivity.py} in the same repository.

\bibliographystyle{apsrev4-2}
\bibliography{references}

\end{document}